\documentclass[preprint,amsmath,amssymb,superscriptaddress, array]{revtex4-1}
\usepackage[graphicx]{realboxes}
\usepackage{dcolumn}
\usepackage{bm}
\usepackage{hyperref}
\usepackage{rotating}
\usepackage[utf8x]{inputenc} 
\usepackage{fancyhdr}
\usepackage{array}
\begin{document}

\title{Tailored Molybdenum Carbide Properties and Graphitic Nano Layer Formation by Plasma and Ion Energy Control during Plasma Enhanced ALD}

\author{Eldad Grady}
\author{Marcel Verheijen}
\author{Tahsin Faraz}
\author{Saurabh Karwal}
\author{W.M.M. Kessels}
\author{Ageeth A. Bol}
\affiliation{Department of Applied Physics, Eindhoven University of Technology, Den Dolech 2, P.O. Box 513, 5600 MB Eindhoven, The Netherlands\\}

\begin{abstract}
We demonstrate the extensive study on how film density and crystallinity of molybdenum carbide ($MoC_{x}$) can be tailored during plasma-enhanced ALD (PEALD) by controlling either the plasma exposure time or the ion energy. We investigated $MoC_{x}$ films grown using $Mo(^tBuN)_2(NMe_2)_2$ as the precursor and $H_2/Ar$ plasma as the co-reactant at temperatures between 150°C and 300°C. We discover a threshold for graphitic layer formation at high mean ion energies during the PEALD cycle. The supplied high energy dose allows for hybridised $sp^{2}$ carbon bonds formation, similar to high temperature annealing. The graphitisation of the $MoC_{x}$ surface takes place at temperature of 300$^{\circ}C$. The graphitic film show a (101) plane diffraction peak with dominant intensity in XRD, and a typical $sp^{2}$ C1s peak along with carbidic metal in XPS measurements. Surface roughness of the film lowers significantly at the graphitisation regime of deposition. This low temperature graphitisation by high energy plasma ions during PEALD shows a great promise to advancing graphene and graphite composites at low temperature by PEALD for future applications.
\end{abstract}
\maketitle
\fancyhead[]{Preprint}
\section{Introduction}

Graphene and graphitic based composites are of great interest in research of next generation electronics, protective layer coatings, and flexible electronics \cite{wu2017growth,guardia2018development}. Transition metal of group IV-VI have shown to be highly suitable for the high temperature CVD graphene growth process due to its thermal stability, and low thermal coefficient matching Si substrates \cite{zou2014carbide} These characteristics result in lower graphene film stress, as the mismatch between graphene and underlying catalytic substrate is significantly reduced.
molybdenum carbide ($MoC_{x}$) is a refractory metal carbide compound that combines the physical properties of ceramics and the electrical properties of metals, with hardness and mechanical strength (300 - 535 GPa), high thermal stability, chemical inertness and a metallic like electrical conductivity \cite{Handbook,HALIM2016406}. The atomic radii ratio of carbon to molybdenum of 0.556 makes it highly suitable to form interstitial carbides with a mixture of covalent, ionic and  metallic bonds between the metal and the carbon atoms which are responsible for its unique set of properties \cite{Handbook}. These characteristics are of great interest for diffusion barriers in ICs, superconductors and various MEMS applications. Furthermore, molybdenum carbide has made a new addition to the 2D material family named MXenes \cite{2dMXene}. As an IC diffusion barrier $MoC_{x}$ is ideally amorphous and dense, while for superconductivity the cubic $\delta-MoC_{0.75}$ shows the highest transition temperature (14.3 K) \cite{Superconductivity_MoC_75} for $MoC_{x}$ films.
 
In order to accommodate the broad spectrum of usage, the ability to separately control the crystallinity and density of $MoC_{x}$ would be an asset for tailoring film properties to the specific application needs.
Thus far, various techniques have been implemented to synthesize cubic $MoC_{x}$, such as high energy ion bombardment of amorphous MoC, CVD, PVD and ultra high pressure synthesis, at high temperatures \cite{1961Natur.191.1194C,lee1987molybdenum,okuyama1985growth}.  While some techniques were successful in the synthesis, the high temperatures required make it unsuitable for integration in temperature sensitive devices. 

Atomic layer deposition (ALD) offers a cyclic soft deposition technique with a broad and low temperature window, submonolayer thickness control due to its self limiting nature, along with high uniformity and conformality. These merits allow for deposition on temperature sensitive applications as well as for deposition of layers sensitive for the kinetic ion impact during sputtering. 

While substantial work has been done on various molybdenum compounds such as molybdenum oxide \cite{MartijnF} and nitride , so far, very little work has been done on ALD of $MoC_{x}$. Recently, a PEALD process using $Mo(^tBuN)_2(NMe_2)_2$ precursor  with a $H_2/N_2$ plasma as co-reactant was developed with the aim of achieving low resistivity films \cite{Bertuch}. Bertuch et al. demonstrated the deposition of a range of $MoC_x-MoC_xN_y$ compounds depending on the  $H_2/N_2$ ratios used. A predominantly molybdenum carbide material with nitrogen impurities was produced with pure $H_2$ plasma at 150 $^\circ$C, which showed the lowest (170 $\mu{\Omega-cm}$) resistivity.
Here, we present the first extensive study on how film density and crystallinity of $MoC_{x}$ can be tailored independently during plasma-enhanced ALD (PEALD) by controlling either the plasma exposure time or the ion energy \cite{profijt2013substrate,faraz2018energetic,faraz2018tuning}. We investigated $MoC_{x}$ films grown using $(^tBuN)_2(NMe_2)_2Mo$ as the precursor and $H_2/Ar$ plasma as the co-reactant at temperatures between 150$^\circ$C and 300$^\circ$C. 
Additionally,  we present the effects of enhancing the impinging ion energy on the physical and chemical properties of $MoC_{x}$ thin films, controlled via RF biasing of the substrate table during the plasma step of the PEALD. 
We discover a threshold for graphitic layer formation at high mean ion energies during the PEALD cycle. While plasma assisted graphene fabrication has attracted significant interest due to the low temperature synthesis \cite{C5NR06537B}, to the best of our knowledge this is the first report of nano graphitic layers formation at low temperature using PEALD.

\section{Experimental methods}

$MoC_{x}$ thin films have been deposited by plasma enhanced atomic layer deposition (PEALD)  at various temperatures and plasma conditions.
In this part the deposition process is explained and the analysis techniques described.

A. Film depositions

Plasma enhanced atomic layer deposition was performed on 100 mm Si (100) wafers coated with 450 nm of thermally grown $SiO_2$.  The depositions were performed in an Oxford instruments FlexAL2 ALD reactor, which is equipped with an inductively coupled remote RF plasma (ICP) source (13.56 MHz) with alumina dielectric tube. The reactor was pumped down to a base pressure of $1\cdot10^{-6 } Torr$ with a turbomolecular pump. The samples were loaded and unloaded from a low pressure loadlock chamber, allowing for a cooldown after deposition in vacuum conditions. The reactor’s deposition table was set to temperatures between 150$^\circ$C and 300$^\circ$C, and the wall temperature to 150$^\circ$C. 

One ALD cycle consists of subsequently a precursor dose step, a purge step, a plasma exposure step and again a purge step. The ($Mo(^tBuN)_2(NMe_2)_2$ precursor ($98\%$, Strem Chemicals) was stored in a stainless steel container, which was kept at 50 $^\circ$C. The precursor container was bubbled by an argon flow of 50 sccm during the precursor dose step, while flowing argon at the same rate from the ICP chamber to avoid deposition in the ICP tube. The total pressure in the reactor during the 6 seconds long precursor injection was set to 200 mTorr. During the plasma exposure step, the ICP RF power was set to 100W. $H_2/Ar$ with 4:1 ratio and total flow rate of 50 sccm was fed from the top ICP tube and the plasma was ignited for the desired exposure time, typically 20 to 80 seconds. Automatic pressure control (APC) valves were fully opened during plasma exposure to reduce the pressure to  7 mTorr. Purge steps (4 to 5 seconds) were set after precursor and plasma half cycles to evacuate residual precursor gas, reaction byproducts and plasma species with 100 sccm argon flow and open APC valves, at pressure 25 mTorr.

B. Film analyses 

Film thickness and optical properties of the deposited films have been studied with a J.A. Woollam UV-spectroscopic ellipsometer (SE) using a model comprising one Drude and two Tauc-Lorenz oscillators. Data was obtained within the range of 190 nm – 990 nm, and refractive index (n) and extinction coefficient (k) were determined. Growth per cycle (GPC) was calculated using in-situ J.A. Woollam IR-SE  every 10 deposition cycles. Optical resistivity was derived from the first term of the Drude oscillator, corresponding to the imaginary part $\epsilon{_2}$.

Sheet resistance values of the $MoC_x$ films deposited on 450 nm $SiO_2$/Si were determined using a Keithley 2400 SourceMeter and a Signatron probe. Film thickness was evaluated using ex-situ UV-SE, to calculate the electrical resistivity at room temperature.

The  film composition was analysed by X-ray photoelectron spectroscopy (XPS) with a Thermo Scientific KA1066 spectrometer, using monochromatic Al $K\alpha$ x-rays with an energy of 1486.6 eV. The films have been sputtered with $Ar+$ ion gun prior to scans, in order to remove surface oxide and adventitious carbon. A continuous electron flood gun was employed during measurements to compensate for charging. XPS studies have been performed to evaluate the film composition as function of plasma conditions. Three main components were evaluated by their corresponding peaks, namely molybdenum by the mo3d peak, oxygen by the O1s peak and carbon by the C1s peak. The ratio between each peak area to the sum of all peaks area gave the partial atomic percentage of the element. The precursor molecule used for deposition contains 4 nitrogen atoms bonded to the molybdenum atom, therefore nitrogen is expected to be found in the film. However, due to overlapping N1s and Mo3p peaks, deconvolution of the N1s peak is not reliable for low N1s peak intensity. 

Rutherford backscattering spectroscopy (RBS) and elastic recoil detection (ERD) were performed by AccTec BV, Eindhoven, The Netherlands, using a Singletron with a 2 MeV He+ beam to  determine the chemical composition of the films. ERD was performed with the detector at a recoil angle of 30$^\circ$. RBS is performed at scattering angles of 150$^\circ$ and 105$^\circ$. The mass density was calculated using the measured mass density as obtained from RBS/ERD measurements and the film thickness as obtained from SE. RBS measurement complement the XPS estimation of film composition, and specifically provide further information on nitrogen content in the film.

Film crystallinity and preferred crystal orientation was studied by Gonio x-ray diffraction. Experiments were conducted with PanAnlytical X'pert PROMRD diffractometer operated using $Cu K\alpha (\lambda=1.54A) $.
A JEOL ARM 200 transmission electron microscope (TEM) at 200 kV was used to analyse the microstructure of films deposited on planar TEM windows. These windows consisted of $\sim15$ nm $Si_{3}N_{4}$ membranes coated with 5 nm of $SiO_2$ grown using ALD. This ensured a $SiO_2$ starting surface while maintaining transparency to the electron beam. High-angle-annular-dark-field TEM (HAADF-TEM) modes was employed to characterise the samples.
\section{Results and discussion}

\begin{figure*}
\centering
\includegraphics[width=0.7\linewidth]{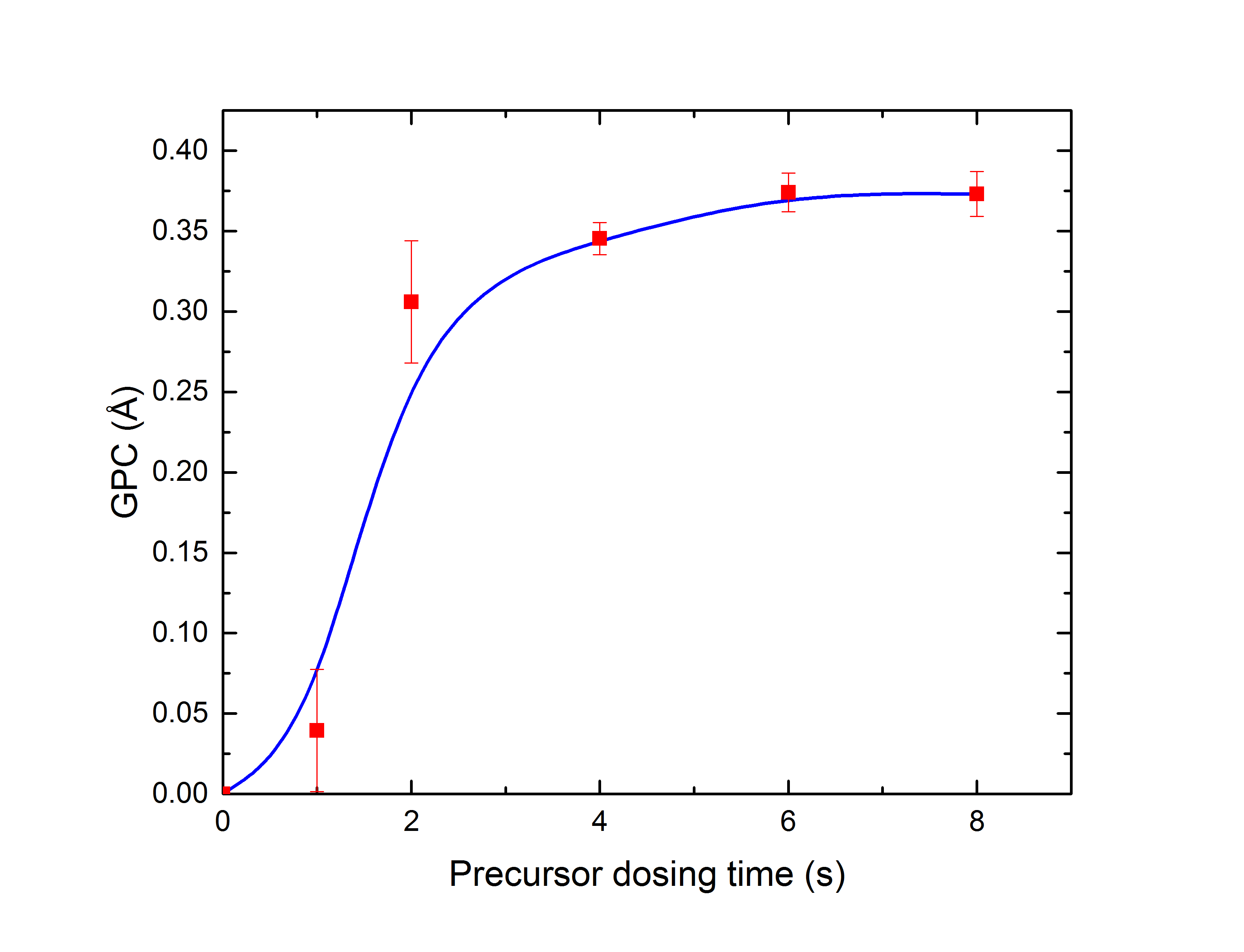}
\includegraphics[width=0.7\linewidth]{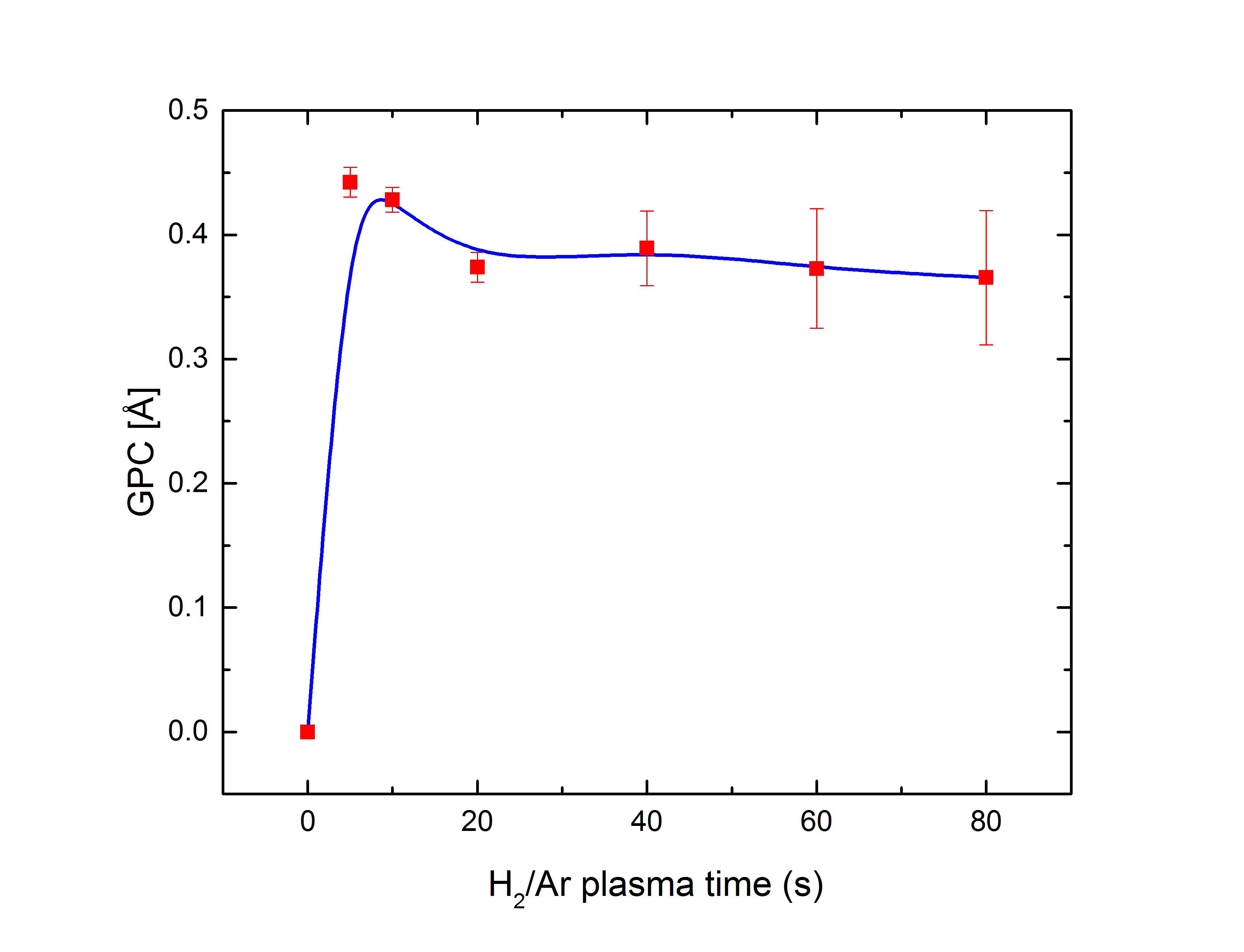}
\caption{Growth per cycle (GPC) of $MoC_{x}$ deposited at 300$^\circ$C determined using a in-situ SE, as a function of (a) precursor dose time (b) plasma exposure time. The lines serve as a guide to the eye}
\label{fig:GPC_precursor_time}
\end{figure*}

\begin{figure}
\centering
\includegraphics[width=0.7\linewidth]{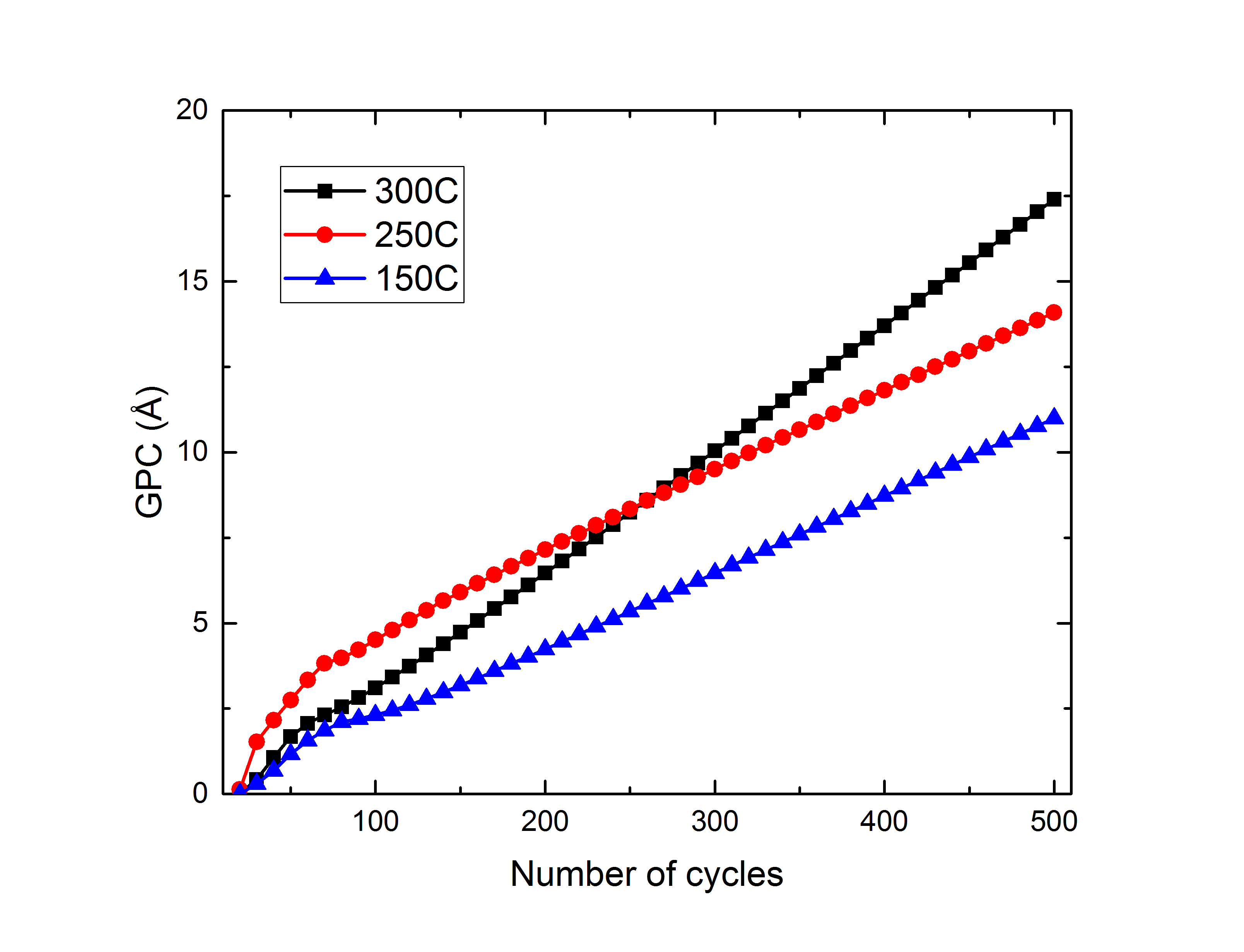}
\caption{Growth per cycle (GPC) during initial 500 cycles. $MoC_{x}$ films deposited using ($Mo(^tBuN)_2(NMe_2)_2$ precursor and$H_2/Ar$ plasma (without substrate biasing) as a function of deposition temperature}
\label{fig:GPC_temp-line}
\end{figure}

Atomic layer film deposition has been performed with $Mo(^tBuN)_2(NMe_2)_2$ precursor and 40/10 sccm of $H_2/Ar$ plasma. The growth per cycle (GPC) was measured with an in-situ SE by measuring the film's thickness every 10 cycles. Precursor saturation occurs after 6 seconds of dosing, and plasma time saturation after 20 seconds of exposure.  The saturation curve for the precursor and the plasma time performed at 300$^\circ$C is presented in figure \ref{fig:GPC_precursor_time}. Typical saturated GPC value is 0.365\AA. The effect of temperature on the GPC in saturated conditions has been studied as well for 150$^\circ$C and 250$^\circ$C yielding  values of about 0.25\AA  (see figure \ref{fig:GPC_temp-line}). The significant increase in GPC at 300$^\circ$C has been explained elsewhere [11] as precursor decomposition. However, we see no evidence to that effect, as the precursor saturation curve confirms no increase in GPC after 6 seconds dosing. Furthermore, film thickness as measured across a wafer with ex-situ SE, show high uniformity typical for ALD. Additionally, previous work in our group using the same precursor for $MoO_{x}$ deposition showed no decomposition at the same  table temperature [12].The characterisation of the $MoC_{x}$ properties with varying temperature provides an insight to observed rise in GPC.

\subsection{Plasma Time and Temperature effects on Films}

\begin{figure}
\centering
\includegraphics[width=0.7\linewidth]{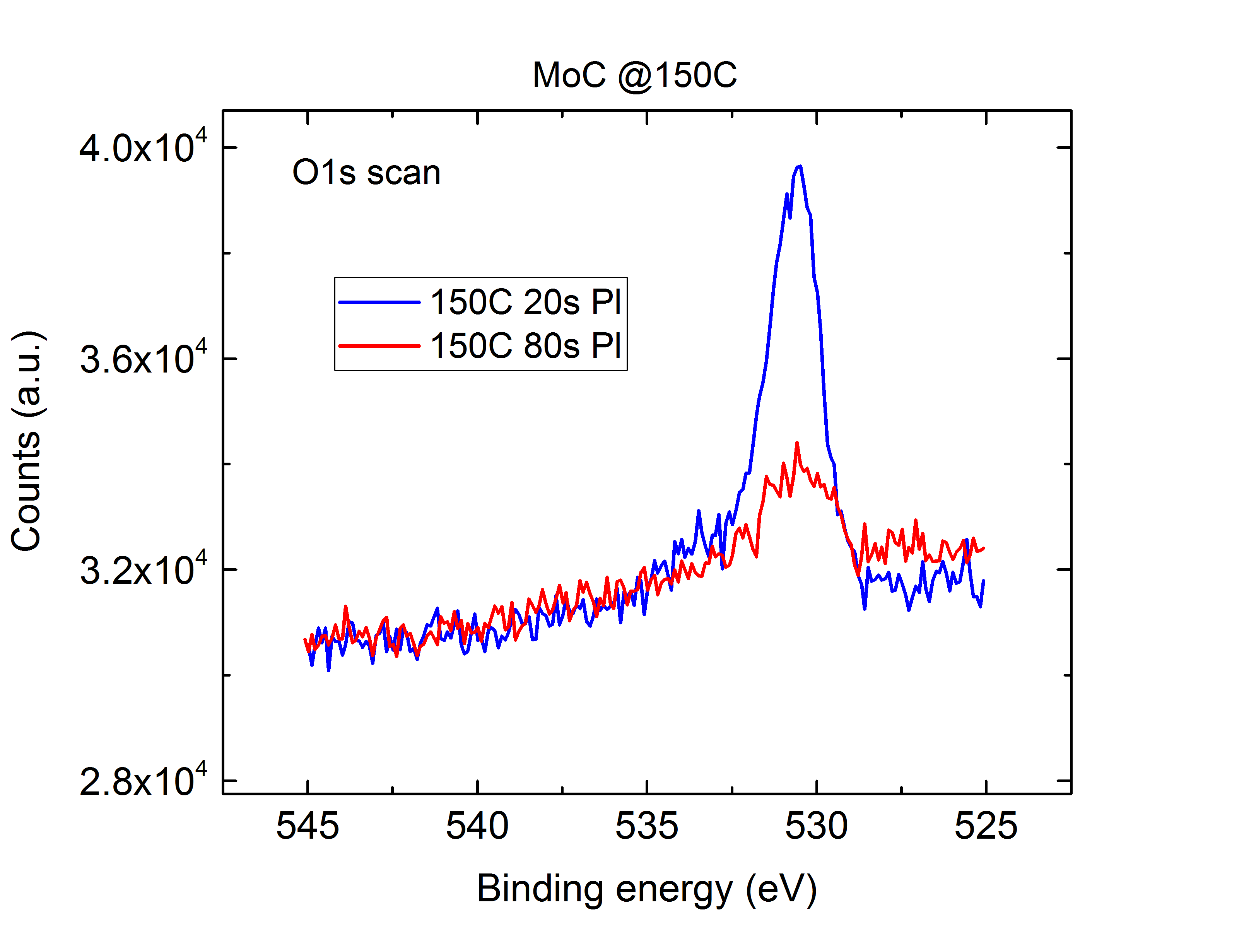}
\includegraphics[width=0.7\linewidth]{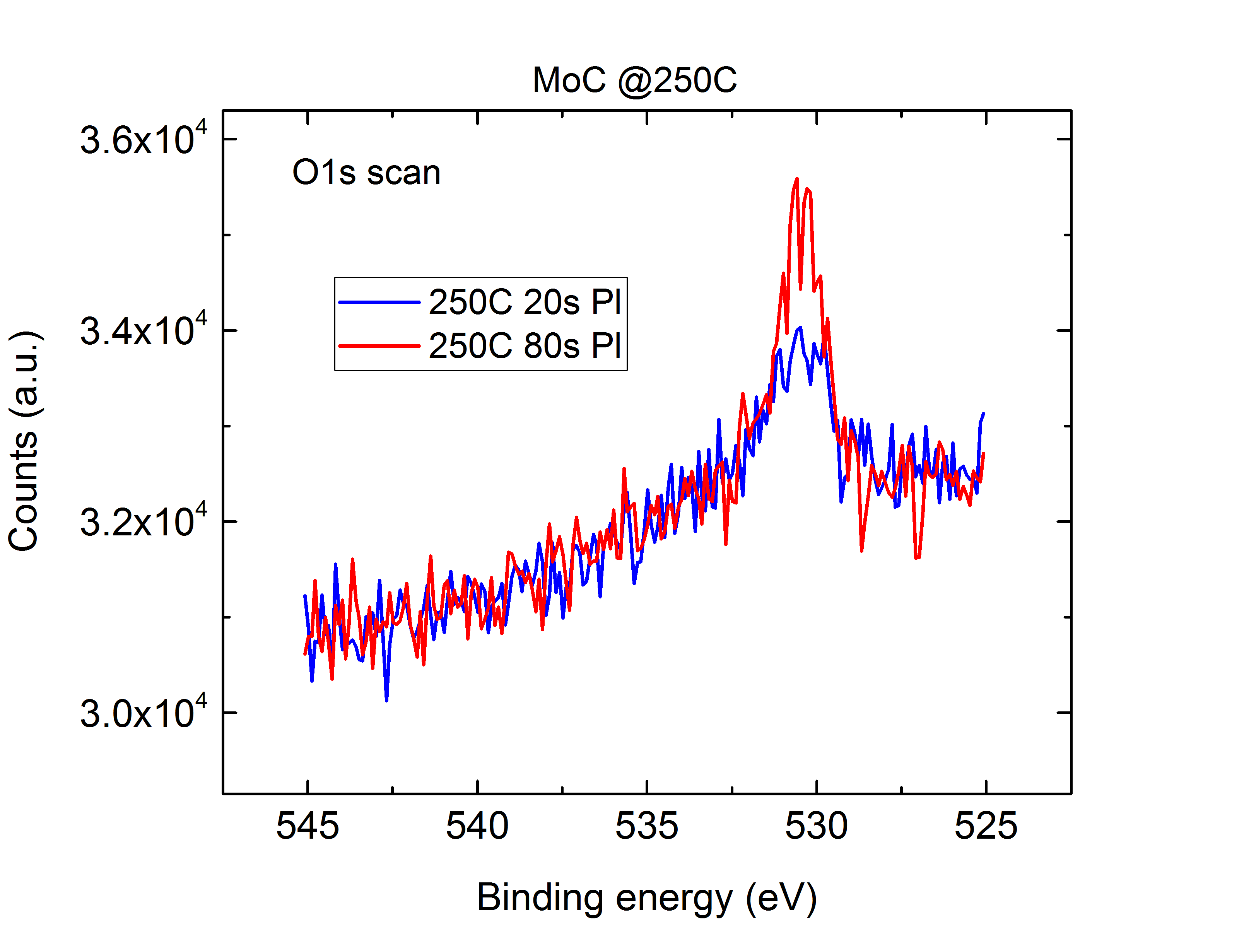}
\caption{XPS core level spectra of the O1s peak for $\sim$22 nm $MoC_{x}$ films deposited at (a) 150 °C and (b) 250 °C. The red and blue lines represent films deposited using $H_2/Ar$ plasma exposure times of 20s and 80s, respectively}
\label{fig:XPS_O1_150C}
\end{figure}

\begin{figure}
\centering
\includegraphics[width=0.7\linewidth]{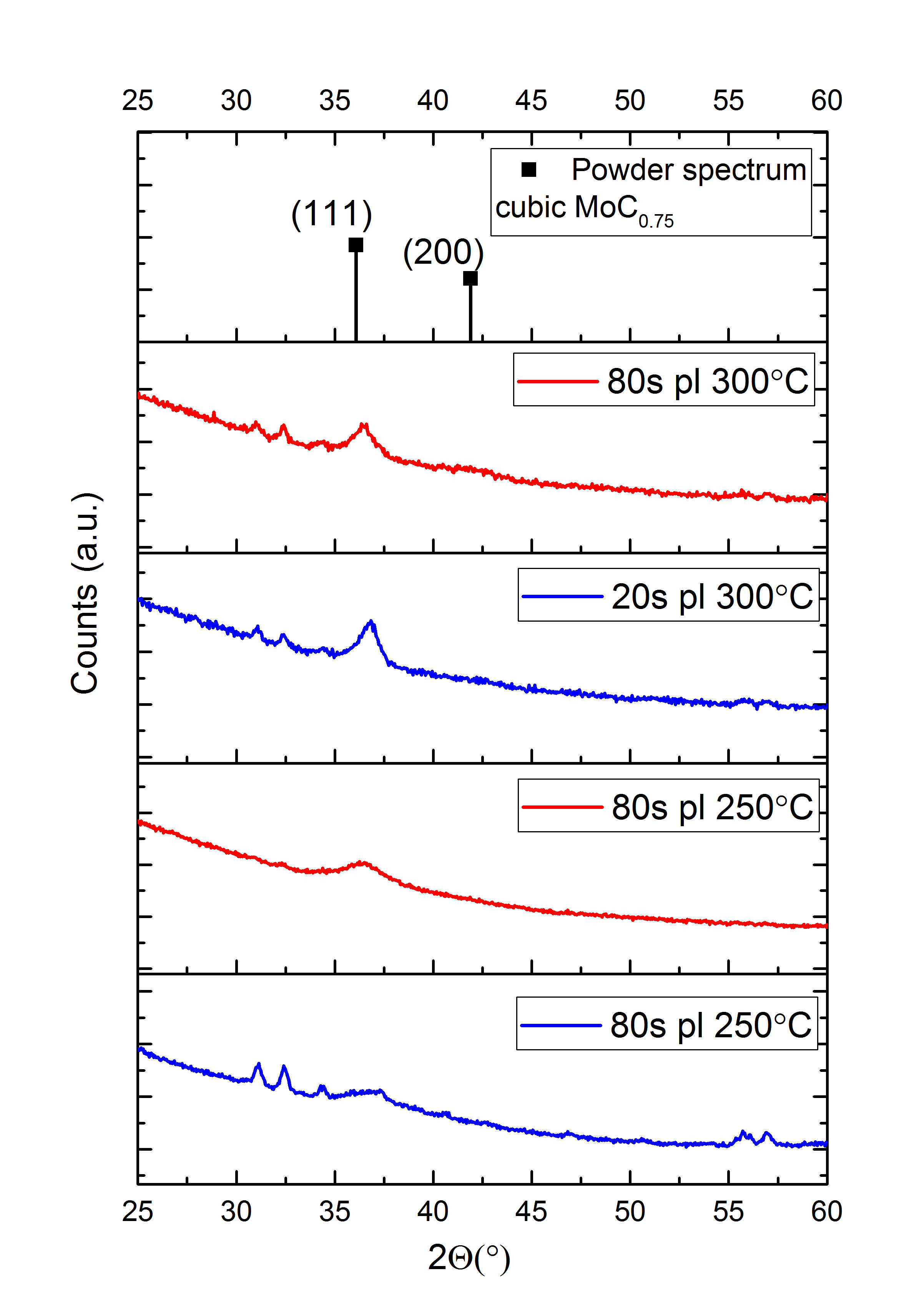}
\caption{$\theta$-$2\theta$ x-ray diffractograms for 30 nm MoC films deposited using different temperatures and $H_2/Ar$ plasma exposure conditions (different plasma exposure times without biasing)}
\label{fig:xrd_temp_plasma_time}
\end{figure}

$MoC_{x}$ depositions at 150$^\circ$C with 20 seconds plasma have produced films with C/Mo ratio of 0.7 as estimated by XPS (see table \ref{fig:tab_150C}). The oxygen content was estimated to be 12.4\% at 150$^\circ$C with 20 seconds plasma. Increasing the plasma exposure time to 80 seconds shows a slight decrease in C/Mo ratio to 0.68 with a sharp decrease of oxygen content to 5.7\%, as demonstrated in figure \ref{fig:XPS_O1_150C}. The growth per cycle (GPC) for both cases remains unchanged at 0.25\AA. All films deposited at 150$^\circ$C show no diffraction peaks in XRD measurements and are deemed to be completely amorphous. Deposition at 250$^\circ$C with 20 seconds plasma produced a film with 5\% hydrogen, 5\% nitrogen and C/Mo ratio of 0.63, however no [O] was detected in the bulk. Hydrogen content could originate both from the plasma coreactant gas, and the [H] rich precursor molecule, while nitrogen content originates solely from the precursor molecule, $C_{12}H_{30}N_{4}Mo$, that contributes 4 [N] atoms. Each of the [N] atoms is directly bonded to the [Mo], and to CH ligands. The $H_2/Ar$ plasma is able to break most N-Mo bonds with 20s plasma, explaining the relatively low rate of [N] in the film. The high content of [C] in the film on the other hand, suggests that the removed ligands are redeposited during the plasma step, ultimately producing a $MoC_{x}$ film. RBS measurements reveal a film mass density of 8.9 $g/{cm^3}$, and [Mo] GPC of 1.33 $\frac{atom}{nm^{2}\times{cycle}}$ and [C] of 0.83 $\frac{atom}{nm^{2}\times{cycle}}$. The XRD pattern show diffraction peaks corresponding to the (111) and (220) lattice planes  of the delta-cubic $MoC_{x}$. The (111) peak width appears broad and distorted, along with several sharper smaller peaks between 28-33$^\circ\, 2\theta$. With 80 seconds plasma at 250$^\circ$C the C/Mo ratio reduced from 0.63 to 0.56 and the nitrogen and hydrogen content from 5\% to below detection limits ($\pm{1}\%$ and $\pm{5}\%$ respectively), as table \ref{fig:RBS1} shows. The removal of [N] content supports the role of the $H_2/Ar$ plasma in removal of N-Mo bonds, and redeposition of [C] from the ligands, allbeit [C] removal effects are revealed with the prolonged plasma exposure. As no [O] was present in the bulk, XPS shows no significant change in O1s peak (see figure \ref{fig:XPS_O1_150C}) with prolonged plasma, as oppose to film deposited at 150$^\circ$C. Moreover, the film's mass density has increased with the longer plasma exposure from 8.9 to 9.2 $g/{cm^3}$. Film deposition at 250$^\circ$C and 80 seconds plasma exhibits a single diffraction peak of the (111) lattice plane, while other diffraction peaks are completely suppressed (see figure \ref{fig:xrd_temp_plasma_time}). The broad peak width indicates small order crystalline domains, embedded in mostly amorphous domains. GPC for both conditions at 250$^\circ$C is at 0.25\AA, similar to deposition at 150$^\circ$C within error margins. The growth of atoms per $nm^{2}cycle$ show a relatively unchanged growth in Mo atoms (1.33 and 1.35 for 20s and 80s respectively), while the growth of C was reduced from 0.83 $\frac{atom}{nm^{2}\times{cycle}}$ to 0.76 $\frac{atom}{nm^{2}\times{cycle}}$ which correlates to the slight increase in density. The removal of [C] content indicates more efficient removal of ligands during deposition with prolonged plasma time. 
Film deposition at 300$^\circ$C with 20 seconds plasma shows a significant increase in GPC from $0.25\AA\,$ for lower temperatures to $0.37\AA$. RBS profiling reveal an increase in C/Mo ratio from 0.63 to 0.92 upon increasing the temperature from 250$^\circ$C to 300$^\circ$C. Atomic growth per cycle a slight increase in Mo (1.35 $\frac{atom}{nm^{2}\times{cycle}}$ at 250$^\circ$C to 1.45$\frac{atom}{nm^{2}\times{cycle}}$ at 300$^\circ$C), while [C] content increased by 60\% from 0.83 to 1.34 $\frac{atom}{nm^{2}\times{cycle}}$. Moreover, 4\% [N] content was detected in the film by RBS and 8\% [H] by ERD measurements. The increase in film impurities content reflected on the mass density, which sharply declined from 8.9 $g/{cm^3}$ to 7.0 $g/{cm^3}$ with the increase in temperature.
 Prolonging the plasma exposure time to 80 seconds results in decreased C/Mo from 0.92 to 0.78, a sharper decrease than that observed at 250$^\circ$C. Correspondently, [Mo] atoms deposition rate increases from 1.45 to 1.63 $\frac{atom}{nm^{2}\times{cycle}}$, and [C] decreases from 1.34 to 1.28 $\frac{atom}{nm^{2}\times{cycle}}$. The increase of [Mo] growth is a factor of 10 higher at 300$^\circ$C than for the same plasma time at 250$^\circ$C ($\Delta{[Mo]}_{250}=0.02\, at./{nm^{2}cycle}$ and $\Delta{[Mo]}_{300}=0.2\, at./{nm^{2}cycle}$). 
Additionally, [N] and [H] content decreased below detection level for 80 seconds plasma. The lower C/Mo ratio and removal of additional impurities reflected in a higher mass density of 8.0 $g/{cm^3}$.
The film deposited with 80 seconds plasma show the dominant diffraction peak at 36.6$^\circ$ $\,2\theta$ of the (111) plane. 
Comparison of film deposited with 80 seconds plasma exposure at different temperatures, shows C/Mo ratio of 0.7 at 150$^\circ$C which is reduced to 0.56  at 250$^\circ$C and then increased to 0.78 at 300$^\circ$C.
By prolonging the plasma exposure time, the total ion energy dose to the deposition substrate is increased linearly with time ($eV_{ion}/s\times{Plasma_{seconds}}$). The increased energy dose resulted in higher mass density and correspondently lower C/Mo ratio and mitigation of film impurities below detection levels. This reflected on the lowered film resistivity, with two fold decrease from 272  $\mu{\Omega-cm}$ to 143  $\mu{\Omega-cm}$ at 300$^\circ$C for 20s and 80s plasma respectively. However, the effect on film crystallinity is not significant.

Generally, the films deposited at 300$^\circ$C show higher peak intensities and narrower peak widths than films at 250$^\circ$C which are mostly amorphous. Hence, more crystalline material is present with larger crystal grains at higher temperature, whose effect on crystallinity is more significant than plasma exposure time. 

\subsection{Biased Substrate effects on ALD Films}

\begin{figure}
\centering
\includegraphics[width=0.7\linewidth]{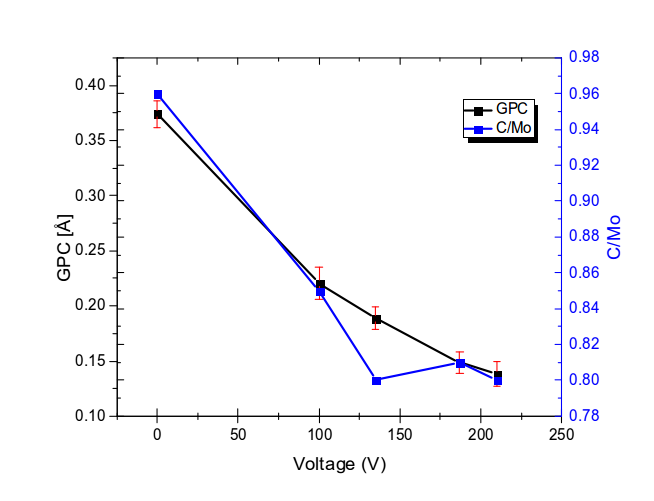}
\caption{Growth per cycle (left axis) and C/Mo ratio (right axis) of $MoC_{x}$ films deposited at 300$^\circ$C as a function of the time-averaged bias voltage applied during the $H_2/Ar$ plasma exposure step.}
\label{fig:XPS_GPC_C_Mo}
\end{figure}

The total energy dose to the deposition substrate can be alternatively altered by increasing the ion energy.
By applying radio-frequency (RF) bias voltage to the substrate we can control the energy of ions impinging on the surface. An RF bias with time average voltage ($V_{bias}$) between -100V to -200V was applied to study the effects on the film properties. For these experiments we use 20 seconds of total plasma time, for a comparison with the saturated non biased plasma time. The plasma during biased ALD is comprised of 10s non biased followed by 10s biased plasma.\newline
The mean energy of the ion energy is calculated from the following equation:\linebreak
$E_{ion} = e \cdot{V_{sheath}} = e \cdot{(V_{plasma} - V_{surface})}$\\
Where $V_{plasma}$ is the ion energy for 0V bias, measured to be 25eV for 100W plasma power at 7mTorr pressure, and $V_{surface}$ is the time averaged bias voltage. Thus for an applied time averaged $V_bias$ ( $<V_{bias}>$) of -100V, the total mean ion energy is 125 eV.
All bias experiments performed were done at 300$^\circ$C, as the largest fluctuation in film density and content occurs at this temperature. 
Figure \ref{fig:XPS_GPC_C_Mo} shows the effect of applied bias on the GPC.
 When $-100V_{bias}$ was applied the GPC declined sharply from 0.37\AA\, at 300$^\circ$C with $<0V_{bias}>$ to 0.22\AA. RBS measurements reveal the cause for this significant decline in a lowered [Mo] and [C] GPC upon applying bias voltage. [Mo] GPC declined from 1.45 to 1.02 $\frac{atom}{nm^{2}\times{cycle}}$ and [C] with over 60\% decline from 1.34 to 0.81 $\frac{atom}{nm^{2}\times{cycle}}$. The corresponding C/Mo ratio reduced from 0.92 for the non biased case to 0.79 with $-100V_{bias}$. An effect of $MoC_{x}$ densification is noted, with mass density measured at 8.2 $g/{cm^3}$ for $-100V_{bias}$ compared to a 7.0 $g/{cm^3}$ for $<0V_{bias}>$.

\begin{figure}
\centering
\includegraphics[width=0.9\linewidth]{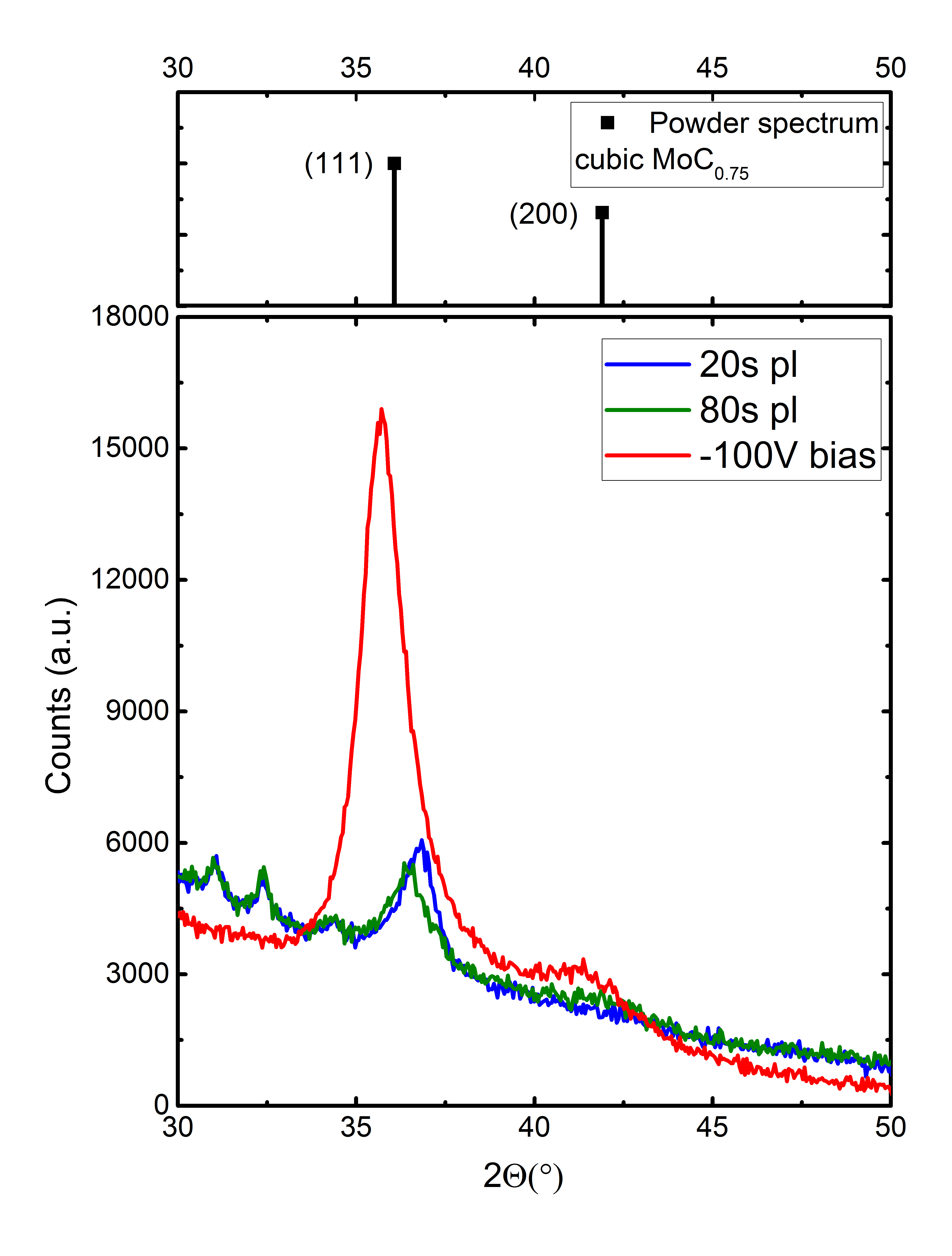}
\caption{$\theta$-$2\theta$ x-ray diffractograms for 30 nm MoCs films deposited at 300$^\circ$C without any substrate biasing with 20 and 80 s plasma, and with a time-averaged bias voltage of -100V applied during the last half (10 s) of the 20 second $H_2/Ar$ plasma exposure step.}
\label{fig:xrd_bias_no_bias_300C_v}
\end{figure}

Physical alterations are also first seen with the application of surface bias voltage, giving the typical cubic $\delta-MoC_{0.75}$  structure with diffraction peaks at 35.75$^\circ$ and 41.42$^\circ$ $\,2\theta$ (see figure \ref{fig:xrd_bias_no_bias_300C_v}).  The dominant peak is the (111) plane, as the (200) plane peak almost completely suppressed. The peak intensity increased by 2 orders of magnitude upon applying $-100V_{bias}$, and peak FWHM decreased by half ($0V_{bias}=1.03\,^\circ2\theta$ and $-100V_{bias}=0.49\,^\circ2\theta$).
Crystallite size has been calculated by fitting the (111) and analysing  the peak broadening using Scherrer equation:
 
 $\tau=K\cdot\lambda/(\beta\cdot\cos{\theta})$     
 
\begin{figure}
	\centering
	\includegraphics[width=0.9\linewidth]{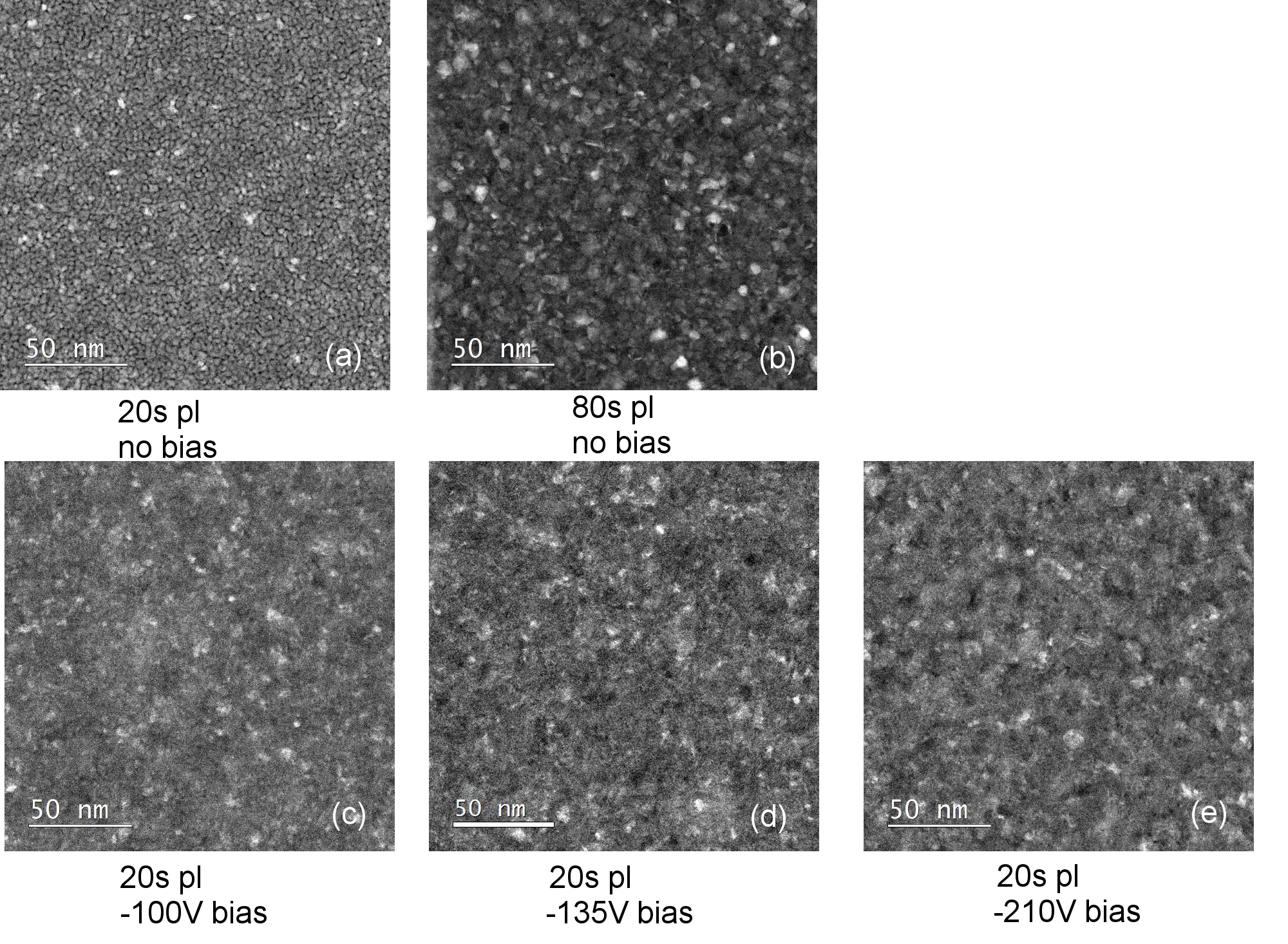}
	\caption{Plan-view high angle annular dark-field (HAADF) TEM images for molybdenum carbide films deposited at 300C with (a) 0 V or no bias, 20 s plasma and (b) 0 V 80 s plasma, and (c) -100 V (d) -135 V and (e) -210 V bias applied during the last half (10 s) of the 20 s $H_2/Ar$ plasma exposure step.}
	\label{fig:TEM_50nm_overview}
\end{figure}

 The XRD data points to a highly crystalline material with crystallite size doubling in comparison for non biased deposition. Crystallite size for 20s pl without bias is calculated to be 9.0 nm while extending the plasma time yields a crystallite size of 7.6 nm, however this slight decrease could be attributed to increased strain in the denser film with 80s pl. When $-100V_{bias}$ bias is applied, crystallite doubles in size to 19.4 nm.
As can be seen in figure \ref{fig:TEM_50nm_overview}, the plan view high angle annular dark-field (HAADF) TEM images support these findings. The size of the visible structured material is significantly increased and voids previously seen for non biased deposition not detectable with $-100V_{bias}$.

\subsection{Graphitisation during PEALD}
With increased $V_{bias}$, further physical alterations and formation of graphitic nano layers are revealed in the $MoC_{x}$ film. With $<-135V_{bias}>$ applied, the (111) lattice peak of the cubic $\delta-MoC_{0.75}$ is diminished, and new diffraction peak appears corresponding to crystalline graphite. The emerging peak at 44.5$\,2\theta$$^\circ$ match the (101) graphite lattice plane and is the dominating peak intensity, indicating strong graphitization with higher ion energies ($E_{ion}=155eV$). The observed emerging phase is consistent with increasing ion energy further ($-187V_{bias},\,E_{ion}=212eV$), however the (101) peak intensity is somewhat diminished in comparison to the previous bias step at $-130V_{bias}$. 
The GPC drops continuesly with increasing $V_{bias}$, from 0.22\AA\, for -100V to 0.15\AA\, and 0.14\AA\, for -135V and -187V respectively. While C/Mo ratio dropped with the application of $-100V_{bias}$ from 0.92 for non biased ALD to 0.79, this ratio has stagnated for higher applied $V_{bias}$, and the C/Mo ratio remained at $\sim$0.80.

\subsection{Plasma Time and Bias Effects on Film Resistivity}
\begin{figure}
\centering
\includegraphics[width=0.7\linewidth]{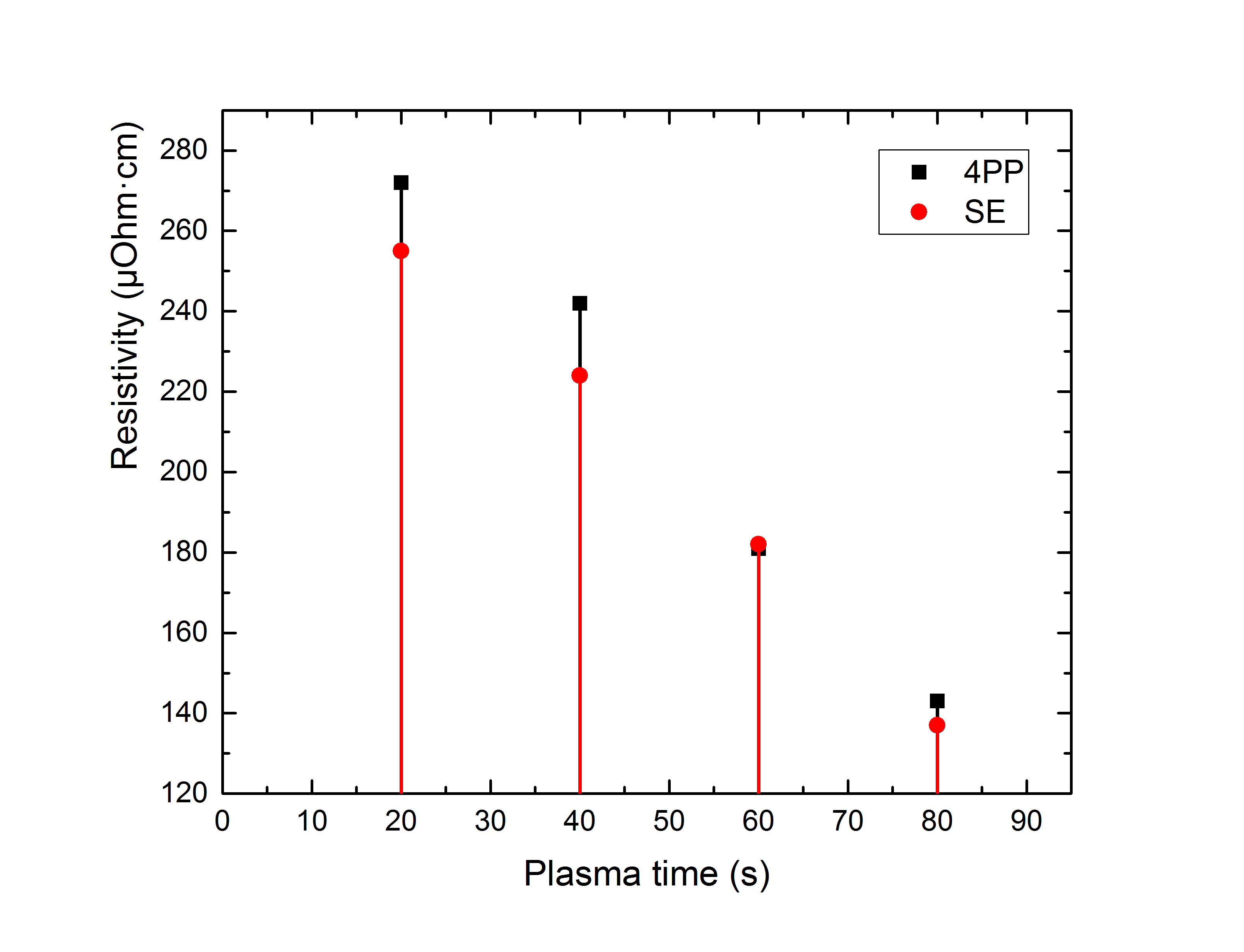}
\caption{Resistivity of $\sim{30}$ nm $MoC_{x}$ films deposited at 300$^\circ$C expressed as a function of the$H_2/Ar$ plasma exposure time. The resistivities were measured using four point probe.}
\label{fig:resistivity_plasma_time}
\end{figure}

The effect of plasma time on $MoC_{x}$ film resistivity has been studied using four-point probe (4PP) measurements. Film of comparable thickness of 30 nm have been deposited at 300$^\circ$C with plasma time ranging from 20 seconds (begin of saturation) up to 80 seconds. Difference in film thickness is within SE error margin. As shown in figure \ref{fig:resistivity_plasma_time}, a significant decrease in resistivity was measured with increase of plasma time. For 20 seconds plasma resistivity was 272 $\mu{\Omega-cm}$ and a decrease to 242 $\mu{\Omega-cm}$  is seen with double the plasma time (40 seconds). Further increase of plasma time to 60 seconds decreased resistivity to 181 $\mu{\Omega-cm}$, and at the final plasma time of 80 seconds 143 $\mu{\Omega-cm}$ was measured. By increasing plasma time from 20 s to 40 s, XPS measurements revealed a reduction of C/Mo ratio from 0.96 to 0.85. With a total 80 s plasma time, the C/Mo ratio decreased further to 0.79. As was demonstrated above, the reduction in C/Mo ratio increased mass density of the $MoC_{x}$ film, and the higher ratio of [Mo] to [C] improves its electrical conductivity.

\begin{figure}
\centering
\includegraphics[width=0.7\linewidth]{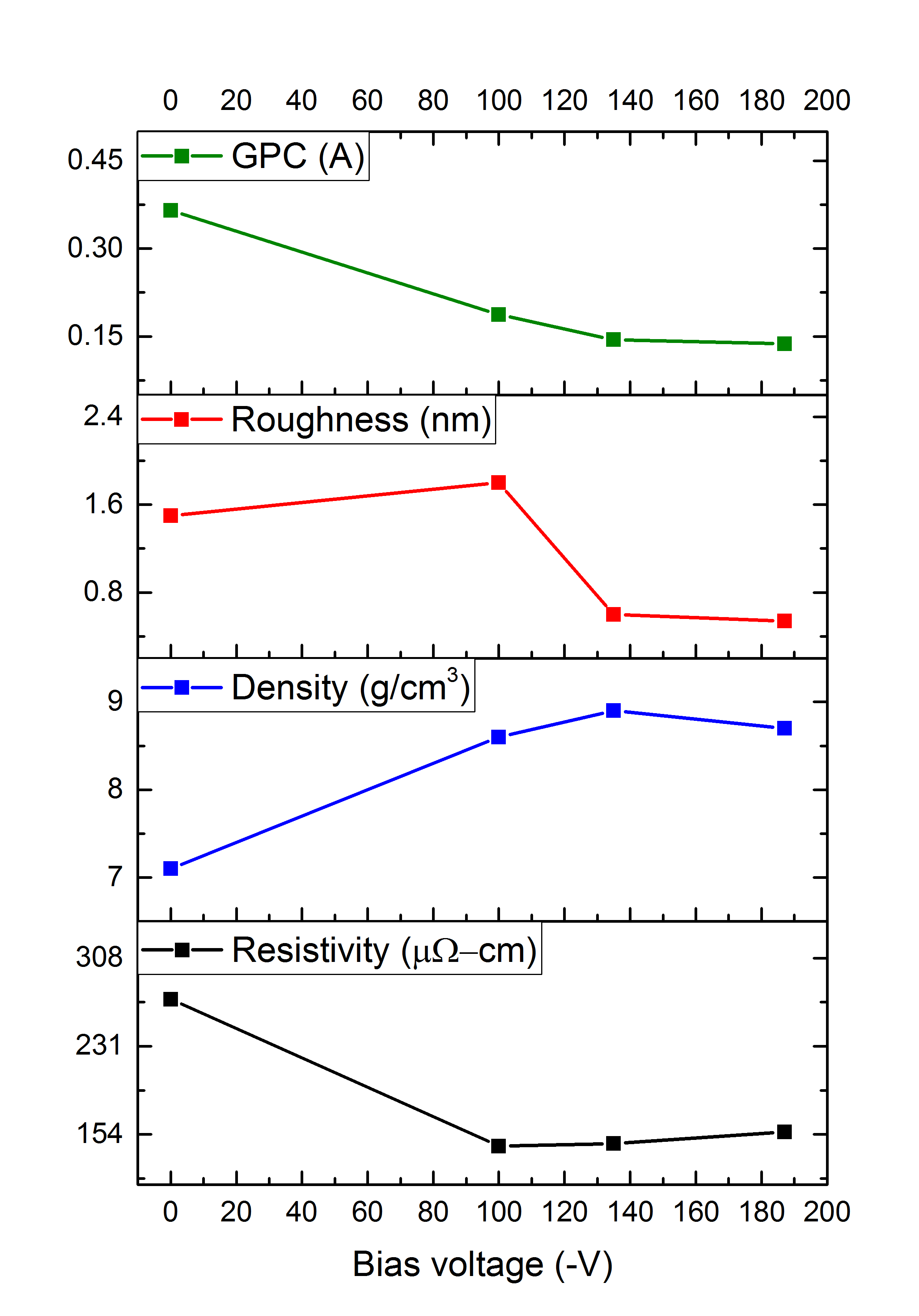}
\caption{Properties of $\sim{30}$ nm $MoC_{x}$ films deposited at 300$^\circ$C with time-averaged bias voltages ranging from 0 V (i.e., no biasing) to 210 V applied during the last 10 seconds of the 20 seconds $H_2/Ar$ plasma exposure step. Graphitisation regime begins with mean ion energies higher than 125eV (-100$V_bias$), and shows a drastic decrease in surface roughness, and a peak in mass density at -135$V_bias$.}
\label{fig:XRR_bias}
\end{figure}

RF bias of the substrate table of average $-100V_{bias}$  has been applied for the duration of 10 seconds, after 10 seconds of non-biased plasma exposure. Film resistivity was measured for films of comparable thickness of 30 nm. We correlate the change in resistivity to GPC and density rates to gain further indications of chemical and physical effects to the film. Figure \ref{fig:XPS_GPC_C_Mo} depicts GPC for biased ALD, and figure \ref{fig:XRR_bias} the resistivity for the corresponding bias voltages. 
With $-100V_{bias}$, film resistivity was decreased to 144 and GPC ratio was decreased from 0.36 (for $0 V_{bias})$) to 0.22.

and 146 $\mu{\Omega-cm}$ respectively. The GPC ratio was decreased from 0.22 \AA\,for -100V to 0.16\AA\, for -135V  $<V_{bias}>$. 
As shown above, the C/Mo reduced upon application of bias voltage from 0.92 to $\sim$ 0.8, which reflected in higher mass density. The produced denser and more metallic film translated to better electrical conductivity.
 Increasing the bias voltage to -187V gives a slight rise in resistivity to 156 $\mu{\Omega-cm}$ and further decrease in GPC to 0.15\AA. At -210V $<V_{bias}>$ a significant increase to 200 $\mu{\Omega-cm}$ is shown with 0.14\AA\, GPC. 

\section{Conclusions}

We presented here the ability to tailor $MoC_{x}$ film by controlling the ion energy and the total energy dose during PEALD. Additionally, we discover graphitic layer formation during the PEALD cycle with mean ion energies between 150eV to 220eV.   $sp^{2}$ hybridised carbon bonds appear to form during the high energy dose to the film surface with average bias voltage is increased to -135$V_bias$. Furthermore, the effects of plasma time exposure, and the ion energy in the plasma during PEALD modification of the $MoC_x$ film electrical, chemical and physical properties were investigated. Both prolonged plasma and biasing are successful in mitigating [N] impurities in the film, and reducing C/Mo ratio. Consequently, a significantly lowered film resistivity was attained of 143  $\mu{\Omega-cm}$. Increasing the total energy dose, which is a product of the mean $E_i$ and the total plasma exposure time. With grounded substrate table ($<0V_{bias}>$) we can increase the total energy dose by prolonging the plasma time, with mean ion energy fixed at 25eV. By increasing the total plasma time from 20s to 80s, we increased film density from 7 to 8, while retaining the film physical properties. With a fixed plasma exposure time of 20 seconds, we increased the mean ion energy by substrate biasing, thus increasing the total energy dose as well. We demonstrated an increase in mass density with increased total energy dose, while no significant change to crystallinity for the same $E_i$. With substrate bias we elevated the impinging ion energies, and demonstrated physical effects to the film. With mean ion energy of 125eV, a highly crystalline and dense film was fabricated corresponding the cubic $\delta-{MoC_{0.79}}$. We can achieve the same high density, low resistivity film in 20s plasma exposure time by biasing, as with long plasma exposure time of 80s without biasing. However, by increasing the mean ion energy, we see physical alterations similar to annealing effects on film, substituting the thermal energy with plasma ion energy. With mean ion energy of $\sim{200}eV$ we see the most crystalline graphitic layer formation, indicating an optimal energy region for graphene formation. The diminishing graphite crystallinity beyond mean ion  energy of 200 eV indicates deterioration of the graphite top layer by high energy $ArH^+$ ions.  The increase in resistivity with further decrease in GPC for $<V_{bias}>$  beyond -135V suggest microstructural changes to the film, more so for the higher -210V. These changes could be in form of voids due to increased ion energy bombardment. The bi-modal distribution of ion energy shows that part of the incident ion have energies higher than -210 eV. The diminished intensity and increased resistivity suggest that these high energy ions at the higher end of the spectrum, could have a sputtering effect on the graphitic top layer when impinging on the surface. Further work needs to be done in optimising the graphene layer by exploring ideal RF bias plasma exposure time, and mean ion energy, as the two determine the total energy dose to the film surface and quality of graphitisation. We postulate, that the high energy ions are able to break Mo-C bonds and achieve a dominantly graphitic film in the [C] rich atmosphere. A trade-off between ion energy and the plasma exposure time needs to be explored, to fine tune the graphitisation.

The applied RF bias voltage effect on film crystallinity and composition is shown to be critical. By controlling the mean ion energy of the impinging ions on the surface of the deposited film, we gain a powerful mean of influence on crystallinity and density alike. These coupled physical and chemical effects are derived from the composition of the coreactant $H_2/Ar$ gas in the plasma. [H] interaction with the surface is mainly chemical in nature, while the heavier Ar atoms contribute to the physical effects with their kinetic energy. The produced $ArH^+$ ions in the plasma combine both these features, hence the coupled chemical and physical effects. By varying the ion energy, we alter predominantly the kinetic impact effect on film, therefore the sharp changes to the $MoC_{x}$ crystallinity.

In conclusion, we demonstrated the control of mass density independent from film crystallinity, and a method to control both features by PEALD, without post deposition thermal annealing treatments, or high temperature deposition.  Furthermore, this work paves the way to fabrication of graphene layers with PEALD, crucial for the production of stacked layers, with atomic precision and high film purity. Moreover, the fabrication can be performed in temperatures as low as 300$^{\circ}C$.  This effect could play a significant role in various applications that require low thermal budget. The ability of control in PEALD of $MoC_{x}$ opens a window for the integration of tailor-made thin $MoC_{x}$, as well as  low temperature graphene films in a wide range of applications previously unavailable.

\subsection*{Acknowledgements}
E. Grady would like to acknowledge the financial support of the Dutch Technology Foundation STW (project number 140930), which is part of the Netherlands Organisation for Scientific Research (NWO).  E. Grady thanks Cristian van Helvoirt for XRD measurements, Jeroen Gerwen and Janneke Zeebregts for their technical support.
\bibliographystyle{unsrt}
\bibliography{MoC_ref.bib}


\begin{table}[p]
	\begin{tabular}{|>{\centering}m {2.5cm}|>{\centering}m {2cm}|>{\centering}m {2cm}|>{\centering}m {2cm}|>{\centering}m {2cm}|}
		\hline  
		\multicolumn{2}{|c|}{} & \multicolumn{2}{|c|} {XPS } & \multicolumn{1}{|c|} {4PP}   \tabularnewline
		\hline
		 \rule[-2ex]{0pt}{5.5ex} Temperature  ($^\circ$C) & Plasma time (s)  & Oxygen (atomic\%)& $\frac{C}{Mo}$ ratio   &
		Resistivity $\mu{\Omega-cm}$ \tabularnewline 
		\hline
		\rule[-2ex]{0pt}{5.5ex}  150 &	20 & 12.4 & 0.70  &	 154   \tabularnewline
		\rule[-2ex]{0pt}{5.5ex}  150 & 80  & 5.6  & 0.72  & 145   \tabularnewline

		\hline 
	\end{tabular} 
	\caption{Atomic \% of elements by XPS and electrical resistivity measurements}
	\label{fig:tab_150C}
\end{table}

\medskip
\begin{sidewaystable}

	\begin{tabular}{|>{\centering}m {2.5cm}|>{\centering}m {2cm}|>{\centering}m {2cm}|>{\centering}m {2cm}|>{\centering}m {2cm}|>{\centering}m {2cm}|>{\centering}m {2cm}|>{\centering}m {2cm}|>{\centering}m {2cm}|}
		\hline  
			\multicolumn{2}{|c|} {} &	\multicolumn{1}{|c|} {ERD} &	\multicolumn{1}{|c|} {XPS} & \multicolumn{5}{|c|} {RBS} \tabularnewline  \cline{3-8}
		\hline \rule[-2ex]{0pt}{5.5ex}Temperature ($^\circ$C) & Plasma time(s) & [H] (atomic\%) &  $\frac{C}{Mo}$ ratio  &  $\frac{C}{Mo}$ ratio   &  [N] (atomic\%) & [Mo] ($\frac{atom}{nm^{2}\times{cycle}}$) &[C] ($\frac{atom}{nm^{2}\times{cycle}}$) & Density ($\frac{g}{cm^3}$) \tabularnewline  \cline{3-8}
			\hline \rule[-2ex]{0pt}{5.5ex} 150 & 20 &  -- &  0.70 &-- & --  & -- & -- & -- \tabularnewline
			\hline \rule[-2ex]{0pt}{5.5ex} 150 & 80 &   --  & 0.72 & -- &  --  & -- & -- & -- \tabularnewline
		\hline \rule[-2ex]{0pt}{5.5ex} 250 & 20 & 5 $\pm{5}$ & 0.64 & 0.63    & 5 $\pm{1}$ &  1.33 &  0.83 & 8.86 \tabularnewline
			\hline \rule[-2ex]{0pt}{5.5ex}  250 & 80  &  $< d.l. $ & 0.75 & 0.56 &  1  &   1.35 &  0.76 & 9.2 \tabularnewline
		\hline \rule[-2ex]{0pt}{5.5ex}  300 & 20  & 8 & 0.96 & 0.92 &  3.5 &  1.45 & 1.34  & 7  \tabularnewline
		\hline \rule[-2ex]{0pt}{5.5ex}  300 & 80 & $< d.l. $ & 0.91 & 0.78  &  $< d.l. $  &  1.63 & 1.28  &   8  \tabularnewline
		\hline 	\multicolumn{8}{|c|} {$< d.l. $: below detection limits of the element. } \tabularnewline
		\hline 
	\end{tabular} 
	\caption{Atomic \% of elements and film density by RBS measurements of non Biased $MoC_{x}$ at various temperatures and plasma times. }
	\label{fig:RBS1}

\end{sidewaystable}
\medskip
\begin{table}[p]
	\begin{tabular}{|>{\centering}m {2.2cm}|>{\centering}m {1.5cm}|>{\centering}m {1.5cm}|>{\centering}m {1.5cm}|>{\centering}m {1.5cm}|>{\centering}m {1.5cm}|>{\centering}m {1.5cm}|>{\centering}m {1.5cm}|}
		\hline  
		\multicolumn{2}{|c|} {} &	\multicolumn{1}{|c|} {ERD} & \multicolumn{5}{|c|} {RBS} \tabularnewline  \cline{3-8}
		\hline \rule[-2ex]{0pt}{5.5ex}Temperature ($^\circ$C) & Plasma time(s)  & [H] (atomic\%) &  [N] (atomic\%) & $\frac{C}{Mo}$ ratio  & [Mo] ($\frac{atom}{nm^{2}\times{cycle}}$) &[C] ($\frac{atom}{nm^{2}\times{cycle}}$) & Density ($\frac{g}{cm^3}$) \tabularnewline 
	\hline \rule[-2ex]{0pt}{5.5ex}    0  & 20 & 8 & 4  &  0.92 & 1.45	& 1.34 &  7.0  \tabularnewline
	\hline \rule[-2ex]{0pt}{5.5ex}    0  & 80 & $< d.l. $ & $< d.l. $  & 0.78 & 1.63 & 1.28 	&    8.0  \tabularnewline
	\hline \rule[-2ex]{0pt}{5.5ex}  -100 & 20 & $< d.l. $ & $< d.l. $    & 0.79 & 1.02 & 0.81  &  8.2  \tabularnewline
		\hline 
		\multicolumn{8}{|c|} {$< d.l. $: below detection limits of the elements. } \tabularnewline
		\hline 
	\end{tabular} 
	\caption{Atomic \% of elements and film density by RBS measurements of non Biased $MoC_{x}$ at various temperatures and plasma times. }
	\label{fig:RBS2}
\end{table}

\medskip


%

\begin{figure}
\centering
\includegraphics[width=0.7\linewidth]{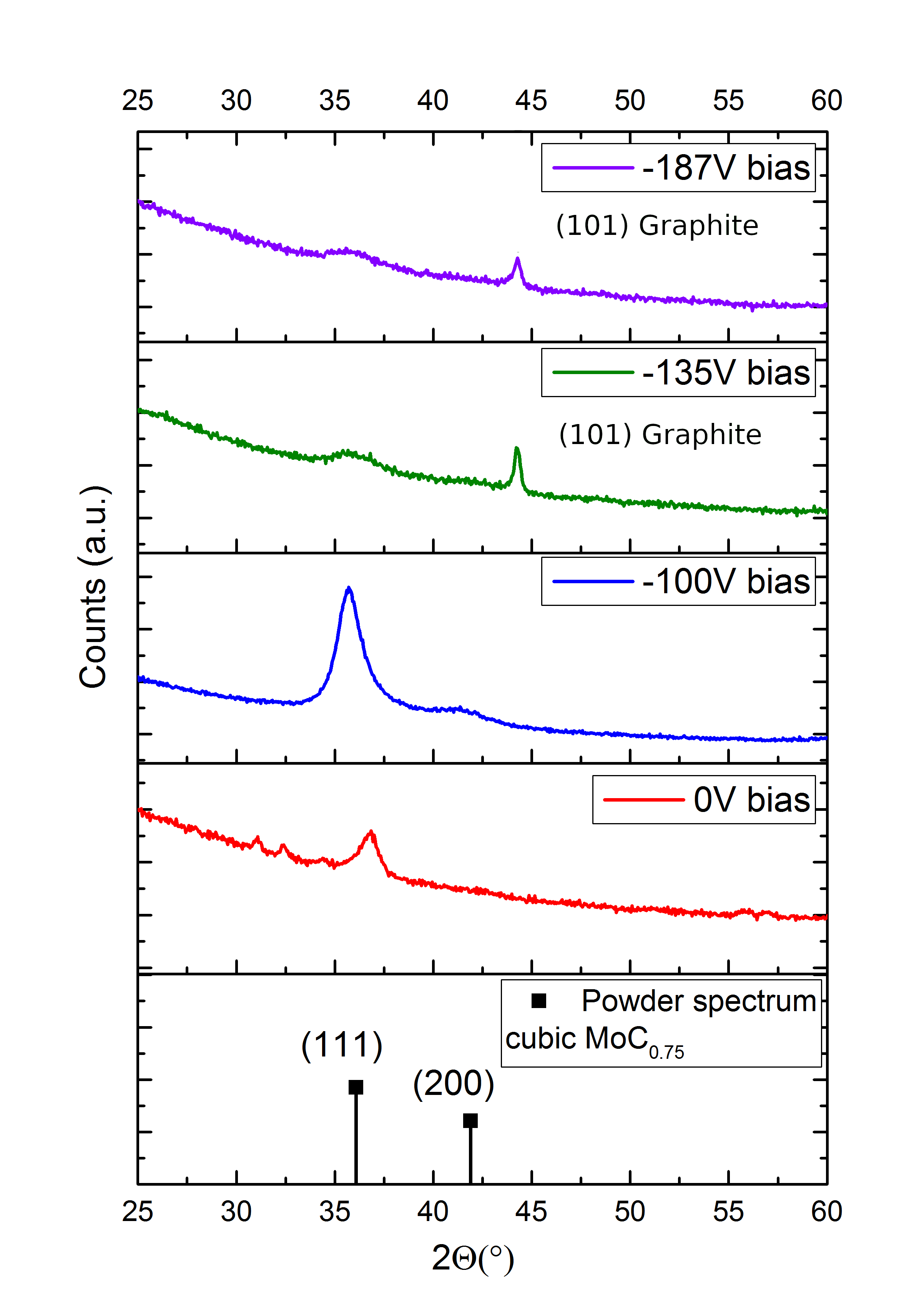}
\caption{$\theta$-$2\theta$ x-ray diffractograms for 30 nm $MoC_{x}$ films deposited at 300$^\circ$C without any substrate biasing with 20, and with a time-averaged bias voltage ranging from -100V to -187V applied during the last half (10 s) of the 20 s $H_2/Ar$ plasma exposure step.}
\label{fig:xrd_bias_v}
\end{figure}

\begin{figure}
	\centering
	\includegraphics[width=0.9\linewidth]{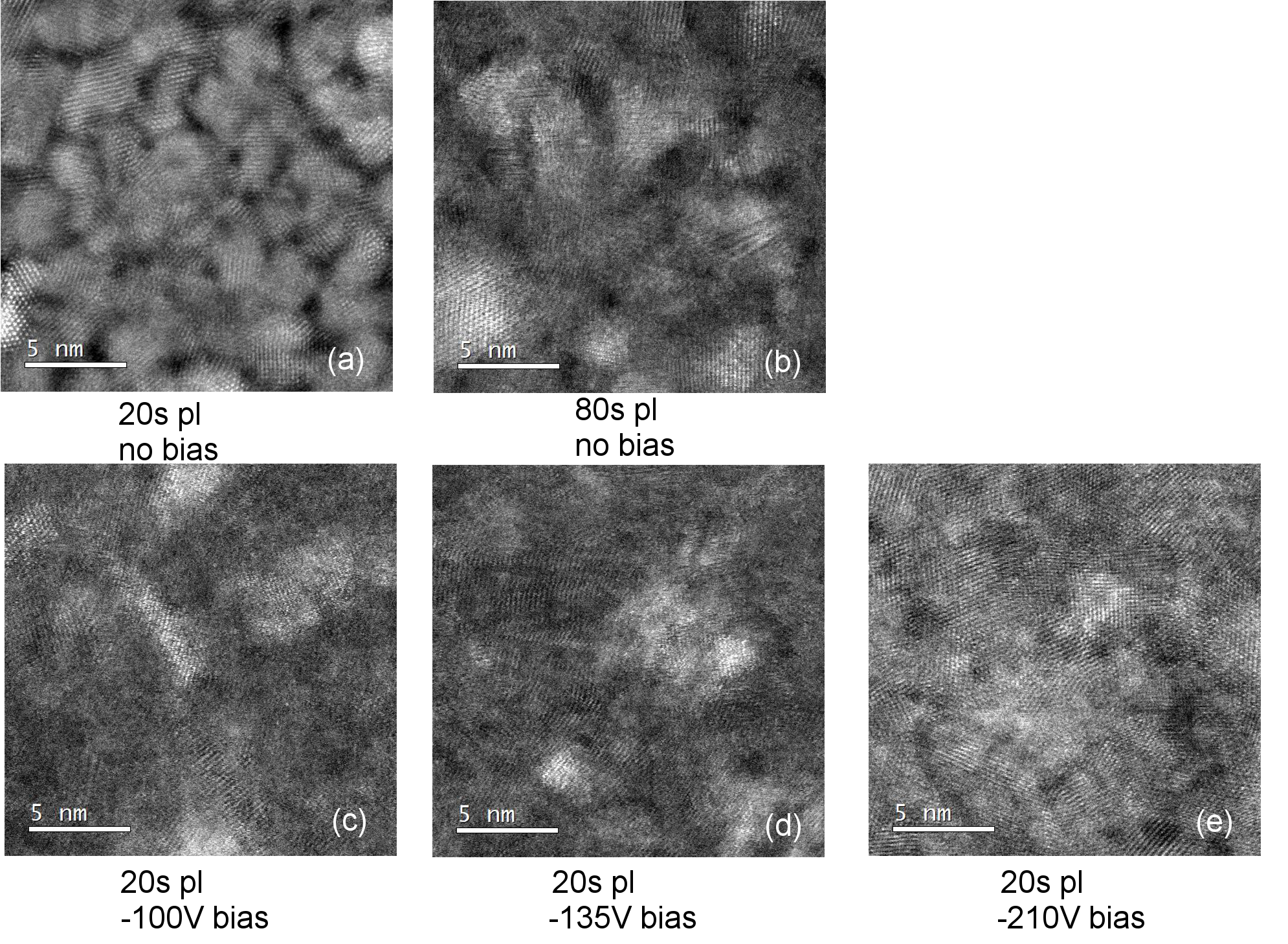}
	\caption{Plan-view high angle annular dark-field (HAADF) TEM images for molybdenum carbide films deposited at 300C with (a) 0 V or no bias, 20 s plasma and (b) 0 V 80 s plasma, and (c) -100 V (d) -135 V and (e) -210 V bias applied during the last half (10 s) of the 20 s $H_2/Ar$ plasma exposure step.}
	\label{fig:TEM_5nm_overview}
\end{figure}

\begin{figure}
	\centering
	\includegraphics[width=0.7\linewidth]{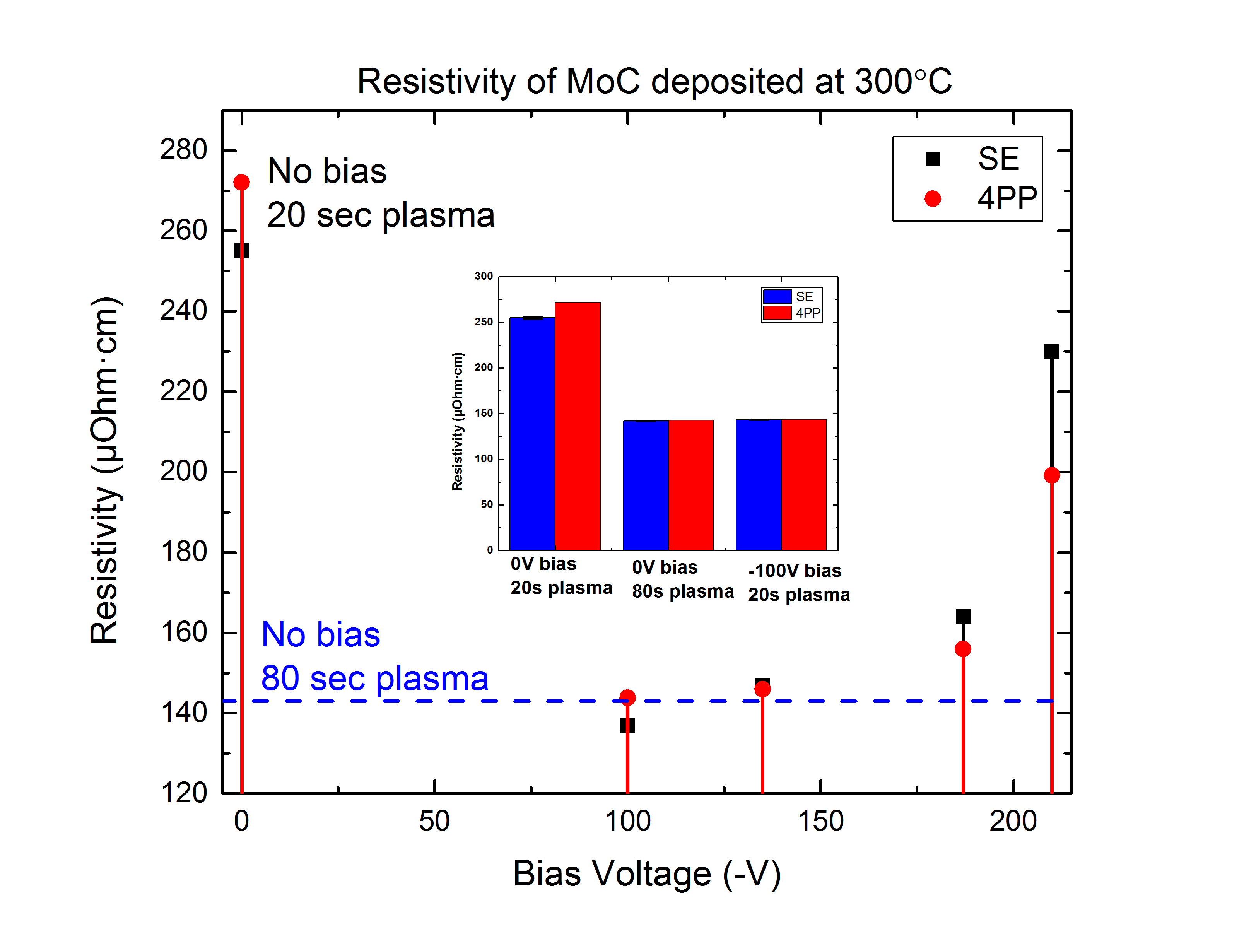}
	\caption{Resistivity of  $\sim{30}$ nm $MoC_{x}$ films deposited at 300$^\circ$C with time-averaged bias voltages ranging from 0 V (i.e., no biasing) to -210 V applied during the last 10 seconds of the 20 seconds $H_2/Ar$ plasma exposure step. The resistivities measured using four point probe are denoted by the stars while those derived using spectroscopic ellipsometry are denoted using squares. Centre: resistivity of $\sim{30}$ nm $MoC_{x}$ films deposited at 300$^\circ$C expressed as a function of the $H_2/Ar$ plasma exposure conditions.}
	\label{fig:resistivity_ALL_bias_no_bias}
\end{figure}
%
\end{document}